# Limitations of gravity models in predicting fine-scale spatial-temporal urban mobility networks

Chia Wei Hsu, Chao Fan, Ali Mostafavi

## Abstract

This study identifies the limitations and underlying characteristics of urban mobility networks that influence the performance of the gravity model. The gravity model is a widely-used approach for estimating and predicting population flows in urban mobility networks. Prior studies have reported very good performance results for the gravity model as they tested it using origin-destination (O-D) data at certain levels of aggregation. Also, the main premise of the gravity model in urban networks is the existence of the scale-free property. The characteristics of urban mobility networks, such as scale-free properties, network size, the existence of hubs and giant components, however, might vary depending on the spatial and temporal resolutions of data based on which these networks are constructed. Hence, the sensitivity of gravity model performance to variation in the level of aggregation of data and the temporal and spatial scale of urban mobility networks needs to be examined. To address this gap, this study examined the basic gravity model, which captures the flow magnitude between O-D pairs based on three variables (population of the origin, population of the destination, and the distance between them). Accordingly, we constructed the urban mobility networks using fine-grained location-based human mobility data for multiple US metropolitan counties. The constructed urban mobility networks have finer resolution as they capture population flow among census tracts on an hourly and daily scale (as opposed to previous studies which used larger spatial blocks). The results show that the scale-free property does not always exist when urban mobility networks are constructed from data with finer spatial and temporal resolution. By examining the association between macroscopic network characteristics, such as the number of nodes and links, average degree, average clustering coefficient, assortativity coefficient, and predictive performance, we found weak association between performance and certain network structures. The findings suggest that: (1) finer-scale urban mobility networks do not demonstrate a scale-free property; (2) the performance of the basic gravity model decays for predicting population flow in the finer-scale urban mobility networks; (3) the variations in population density distribution and mobility network structure and properties across counties do not significantly influence the performance of gravity models. Hence, gravity models may not be suitable for modeling urban mobility networks with daily or hourly aggregation of census tract to census tract movements. The findings highlight the need for new-generation urban mobility network models or machine learning approaches to better predict fine-scale and high temporal-resolution urban mobility networks.

______________________________

Zachry Department of Civil & Environmental Engineering, 3136 TAMU Texas A&M University, College Station, TX 77843-3136

**Corresponding author**:

Chia Wei Hsu, Zachry Department of Civil & Environmental Engineering, Texas A&M University, 3136 TAMU, College Station, Texas

Email: chawei0207@tamu.edu

# Introduction

Urban mobility networks have attracted attention due to their application in enhancing transportation and urban planning efficiency and for predicting epidemic spread by characterizing the flow of human movements (Barbosa et al., 2018; Dong et al., 2020; Fan et al., 2021; Gao et al., 2021; Glaeser et al., 2020; Jensen, 2009; Yan et al., 2017). Network properties and patterns of population flow across urban regions provide valuable insights on specific modeling settings and for deriving plans targeting the requirements of urban regions under unique circumstances (Noulas et al., 2012; Wang et al., 2018). Data sources with varying spatial and temporal resolutions can be used to construct urban mobility networks; therefore, it is critical to understand the sensitivity of urban mobility network models by studying the structure and properties of these networks and using different level of aggregation of population flow data (Sevtsuk and Ratti, 2010).

As mobile devices have become ubiquitous and the positioning technology more advanced, geocoded datasets recording human movement activities have become more precisely representative of human activity and available for study (Calabrese et al., 2010; Jiang et al., 2013). Multi-source data collected from personal devices, GPS, and social media have driven a new era in urban studies that enables modeling of human mobility at a finer scale than that afforded by previous coarser-scale perspectives. (Wang et al., 2019). Prior urban mobility studies primarily employed traffic analysis zones or ZIP codes as a spatial unit of analysis (i.e., nodes in mobility networks) (Population flows across ZIP codes are typically aggregated on a monthly or annual basis (Martínez et al., 2009).) Nowadays, we can acquire the locations, timestamps, and even the trajectories of individuals and aggregate them into finer-grained spatial levels, such as census tracts or census block groups, and at finer temporal scales, such as hours and days (Kong et al., 2018). Various models and metrics have been introduced based on the specific needs of different disciplines and the nature of data sources. One stream of existing studies investigates network science approaches, which describe the structure of the network with nodes and weighted links. The scale-free property in networks was a fundamental discovery in the network science field and has become a common assumption for many real-world networks, such as mobility networks (Barabási and Albert, 1999; Barabási and Posfai, 2016; Newman, 2018). To examine the scale-free property in networks, the power law on degree distributions is employed as a metric of scale-free networks (Liu et al., 2017; Zhang et al., 2019). The existence of the scale-free property in urban mobility networks has been reported by other studies (Barabási, 2009; Zhang et al., 2019), and this finding led to the prominent use of the gravity model for predicting urban mobility (Jung et al., 2008; Pourebrahim et al., 2018; Thompson et al., 2019).

Existing literature suggested that the gravity model performs well on networks that have scale-free properties in multiple research contexts (Anderson, 2011; Tinbergen, 1963). For example, air and public ground transportation and human migratory and international trade networks (De Benedictis and Taglioni, 2010; Greenwood, 2005; Jung et al., 2008; Wojahn, 2001). The model has broad applications related to spatial interactions within the fields of transportation, economics, geography, urban planning, public health, and politics, and the intersections among them. The gravity model usually performs well on coarse-grained data, such as annual or monthly international trade flows between countries or continents; and daily or monthly population flows between countries, cities, transportation analysis zones (TAZs), or regions tagged by ZIP codes. One focus of the literature is enhancement of the predictive performance of the gravity model by introducing domain-specific independent variables and quantifying the effect of each independent variable on the dependent variables (Huff and Jenks, 1968; Mikkonen and Luoma, 1999; van Bergeijk and Brakman, 2010). Despite the acceptable performance of the gravity model in predicting urban mobility networks built upon coarse-grained population flow data, the performance of gravity model in predicting finer spatial and temporal scale mobility has not been tested.

To address this gap, in this study, we investigated the performance of gravity models in predicting urban mobility networks using finer spatial and temporal aggregations. The main research questions are: (1) Does the scale-free property apply to finer-grained urban mobility networks? (2) To what extent gravity model is capable of predicting finer-grained urban mobility networks even when the scale-free property does not exist? and (3) Does the gravity model perform better in predicting mobility networks in certain cities with particular mobility network structures and properties? To answer these research questions, we created origin-destination (O-D) networks in which nodes are census tracts and links are trip counts with both hourly and daily aggregation. Network metrics are used to examine the changes in network structures through time and to examine whether the scale-free property is observed in these finer-grained urban mobility networks. The association between the evolution of network structures and the performance of the gravity model would reveal the circumstances under which the gravity model performs better that coarser-scale perspectives even when the scale-free property does not exist. The findings provide valuable insights for researchers and practitioners working on modeling spatial interactions and may transform the prediction of human mobility in cities.

## Methods

### Data collection and preprocessing

Three primary datasets were used in this study to extract human mobility characteristics and to collect urban demographics. To construct human mobility networks, the X-Mode dataset and US census tract geodata were processed into O-D trip counts at the census-tract level. Population and geodata from the United States Census Bureau (USCB) was further processed to obtain population density for each census tract, which is one of the factors required by the gravity model to predict population flows.

The mobile phone data are provided by X-Mode, a location intelligence company which collects locations of anonymized mobile phone devices whose owners opt-in to share their location information. X-Mode has been collecting the data for 50 million (MM) mobile phone devices globally through General Data Protection Regulation-compliant and California Consumer Privacy Act-compliant framework. More than 30 MM mobile phone users with 2 billion to 3 billion location data every day are collected in the United States. The data was shared under a strict contract with X-Mode through their academic collaborative program, in which they provide access to de-identified and privacy-enhanced mobility data for academic research. All researchers processed and analyzed the data under a non-disclosure agreement and were obligated not to share data further or to attempt to re-identify data. The X-Mode dataset is one of the most comprehensive location-based datasets of anonymized mobile devices. All records, anonymized and de-identified, contain data that includes information about the mobile devices, their locations and timestamps, their heading direction and speed, points of interest at their locations, and information about data sources.

The 2020 US census tract geodata acquired from the UCSB contains the geographic information of each census tract, including ID, boundaries, centroid, and area. From the X-Mode dataset, we assigned each device's stopping point to individual census tracts with the assistance of the US census tract geodata. If a stopping point moved from one census tract to another, we created a link between these two census tracts. Then we counted the number of movements in a time frame. The mobility networks were constructed with nodes and links, where nodes represent census tracts, links represent the connection between census tracts, and the link weights are the number of movements aggregated both hourly and daily. Figure 1 is a sample illustration of the mobility network of Harris County, Texas, and Queens County, New York. (Queens County is a borough of New York City.) Red points are the centroids of census tracts; their size represents the total number of trips starting or ending within the census tract. Black lines are the trips; their thickness represents number of trips. We also used population information from each census tract area to determine population density of each census tract as an input to the gravity model.

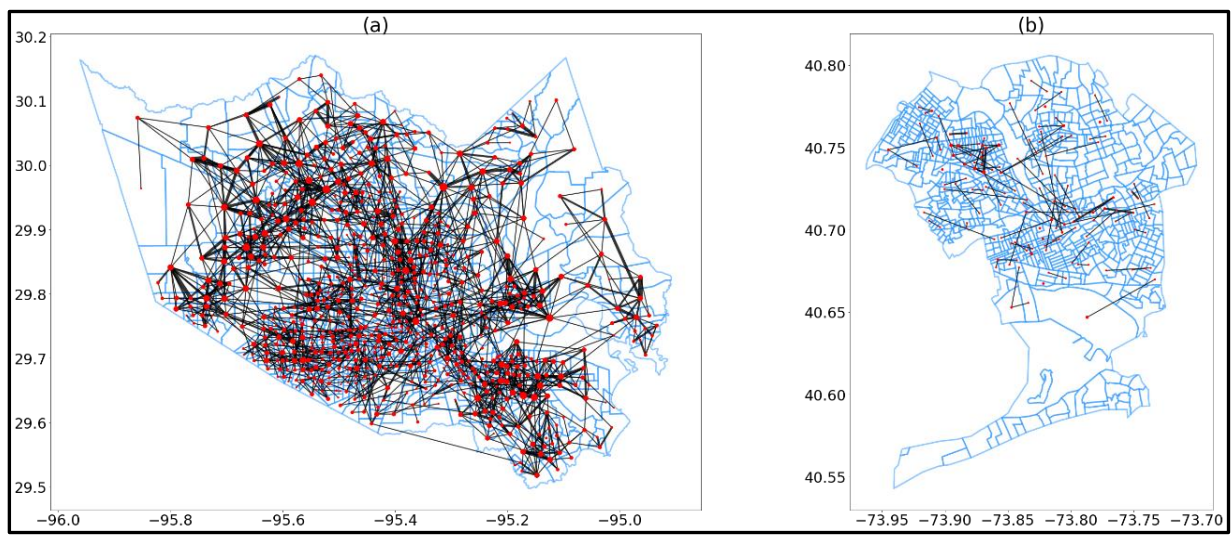

**Figure 1.** Mobility networks at 6:00 p.m., February 1, 2020. (a) Harris County, Texas (b) Queens County, New York.

We aggregated the mobility data of February 2020 into both daily and finer hourly temporal resolution and constructed mobility networks. As February 2020 preceded the outbreak of the COVID-19 pandemic, the data collected during this period represents a normal or control condition without perturbation by any crisis. The spatial resolution of the networks in this study is on the census-tract level (nodes are census tracts) to compare with literature using coarser spatial data resolution (such as transportation analysis zones and ZIP codes). We examined the mobility network in 18 metropolitan counties in the United States, including Bronx, Kings (known as Brooklyn), Queens, and Richmond (known as Staten Island), all boroughs of New York City; Clackamas, Multnomah and Washington of Oregon; Collin, Dallas, Denton, and Harris of Texas; DeKalb and Fulton of Georgia; DuPage of Illinois; Suffolk of Massachusetts, San Francisco of California; Wayne of Michigan. The structural characteristics of the mobility networks vary between counties. Table 1 shows the basic characteristics of mobility networks for each county for February 2020. In the following sections, we will discuss our analysis for testing the performance of the gravity model on these fine-grained urban mobility networks.

**Table 1.** Mobility network characteristics of each county.

| County | Number of Nodes | Number of Edges | Average Weighted Degree | Average Clustering Coefficient | Average Shortest Path Length | Average Assortativity Coefficient |
|---|---|---|---|---|---|---|
| Bronx | 339 | 22918 | 198.18 | 0.00 | 1.74 | -0.04 |
| Clackamas | 80 | 3364 | 696.06 | 0.01 | 1.37 | -0.02 |
| Collin | 152 | 14099 | 1961.72 | 0.00 | 1.28 | 0.00 |
| Dallas | 529 | 71754 | 1020.37 | 0.00 | 1.65 | -0.04 |
| DeKalb | 145 | 8529 | 545.41 | 0.00 | 1.44 | -0.03 |
| Denton | 137 | 9586 | 1647.03 | 0.00 | 1.38 | 0.01 |
| DuPage | 216 | 15014 | 697.92 | 0.00 | 1.58 | -0.05 |
| Fulton | 204 | 14470 | 836.88 | 0.00 | 1.56 | 0.10 |
| Harris | 786 | 129486 | 1383.35 | 0.00 | 1.71 | 0.01 |
| Kings | 760 | 64131 | 230.23 | 0.00 | 1.79 | 0.00 |
| Multnomah | 171 | 11644 | 528.08 | 0.00 | 1.49 | -0.02 |
| New York | 287 | 29663 | 353.42 | 0.01 | 1.60 | -0.02 |
| Queens | 668 | 52269 | 277.52 | 0.00 | 1.82 | -0.04 |
| Richmond | 109 | 7560 | 657.23 | 0.01 | 1.31 | -0.01 |
| Suffolk | 203 | 11328 | 283.33 | 0.00 | 1.65 | -0.07 |
| San Francisco | 195 | 12634 | 269.36 | 0.01 | 1.60 | -0.07 |
| Washington | 104 | 6281 | 807.59 | 0.00 | 1.31 | -0.04 |
| Wayne | 609 | 41068 | 393.67 | 0.00 | 1.85 | 0.08 |

## Fitting the Gravity Model

In this study, the gravity model is applied to predict the flow pattern of human mobility networks across cities under different temporal resolutions. The fitting performance is measured by $R^2$. The gravity model uses the gravitational force concept as an analogy to explain the intensity of spatial interaction, such as the context of human movements in this study, enhanced to incorporate more complicated forms with more variables to facilitate the adaptation of the model to other contexts. As the family of gravity models grows larger with some designed for specific disciplines and data with special attributes, more advanced parameter estimation measures were also suggested. The basic form of the gravity model, however, is still widely applied for its broad interdisciplinary applications for spatial interaction and ability to model urban mobility with coarse-grained datasets as a scale-free model. Therefore, it is important to test the reliability of the basic gravity model if the data has finer temporal and spatial resolution in order to provide a reference about whether more advanced models should be chosen.

The expressions of the basic gravity model are very similar across applications and contexts. The population of origin and destination and the distance of separation between the two are used to estimate the flow (Tinbergen, 1963). The basic assumption is that there will be a positive association between the populations of two regions and the flow, while the level of interaction is expected to be mitigated when the distance separation of the two regions increases. The basic form of the gravity model is shown in Equation 1

$$F_{ij} = G \frac{P_i^{\beta_1} P_j^{\beta_2}}{D_{ij}^{\beta_3}} \tag{1}$$

where, $F_{ij}$ is the flow volume of human movements, $P_i$ and $P_j$ are the population sizes of the origin $i$ and the destination $j$, and $D_{ij}$ is the distance between $i$ and $j$. $\beta_1$, $\beta_2$, and $\beta_3$ are the parameters to be estimated. $G$ is a scaling factor, and flow $F_{ij}$ between $i$ and $j$ is proportional to $\frac{P_i^{\beta_1} P_j^{\beta_2}}{D_{ij}^{\beta_3}}$. A logarithmic operator can be applied to form a log-linear model where $\beta_0$ is the intercept and $\varepsilon_{ij}$ denotes the random disturbance error term as shown in Equation 2

$$logF_{ij} = \beta_0 + \beta_1 logP_i + \beta_2 logP_j - \beta_3 logD_{ij} + \varepsilon_{ij} \tag{2}$$

The error $\varepsilon_{ij}$ accounts for the true error and measurement error. The parameters can then be estimated by ordinary least squares. The values of estimated parameters show how independent variables affect the dependent variable. $R^2$ is one of the important indicators in measuring how close the data are to the fitted model and the fraction of variance explained. Whether an $R^2$ value is considered acceptable or good enough depends on the disciplines and research contexts (Moksony, 1999). For some transportation and economic studies, the $R^2$ value using the basic gravity model can reach 0.7, while the enhanced or adjusted gravity models reported better $R^2$ values(Li et al., 2020; Shen, 2004; Thompson et al., 2019). These reported $R^2$ values give us references for the performance evaluation of basic gravity models in our finer resolution mobility networks in this study.

## Results

### Scale-free property in the mobility networks

Generally, scale-free networks are characterized by the presence of large hubs. Only a few nodes are highly connected to others. Their degree distributions follow power law which can be expressed as shown in Equation 3

$$p_k \sim k^{-\gamma} \tag{3}$$

where, $p_k$ denotes the fraction of nodes with degree $k$, and $\gamma$ is the degree exponent (Barabási, 2009; Barabási and Posfai, 2016). $p_k$ decays slowly as $k$ increases. The degree distribution would take the form a straight line if $p_k$ is plotted against $k$ on a log-log scale. That is, $log\, p_k$ is linearly dependent on $log\, k$. This property is called scale-free because power laws have the same functional form irrespective of network size. The relationship becomes $p_k(ak) = a^{-\gamma} p_k(k)$ when $k$ is rescaled, differing only by a multiplicative factor (Albert and Barabási, 2002).

There is no single clear standard for the criteria to determine the existence of scale-free property. Most studies in the literature utilize the degree exponent from the degree distribution of a network as the main criterion (Barabási and Posfai, 2016). Some studies suggest that a network is scale-free when the degree exponent of a network is larger than 1, while some literature has stricter definitions, requiring the degree exponent to be between 2 to 3 to be categorized as scale-free (Broido and Clauset, 2019).

Figure 2 shows the distribution of degree exponents for degree, in-degree, and out-degree for the mobility network of Harris County in February 2020. The degree exponents are obtained from the slope of the regression line of the degree distribution on a log-log scale.

For hourly networks, the $R^2$ of the linear fitting is around 0.7, which indicates a satisfying fitting result. We can also observe that the distribution follows the power law by the loosely defined scale-free property, since most of the exponents are larger than and suggests that the distribution follows the power law by the loosely defined scale-free property since most of the degree exponents are larger than 1. The mobility network gets close to a scale-free form during rush hours; however, scale-free property is not universal among the hourly mobility networks if the stricter definition of scale-free property is applied. The patterns of fluctuation are similar for degree, in-degree and out-degrees. Our results are consistent with the studies that pointed out that networks with a strong scale-free property are rare in the real world. This raises a question about whether the scale-free assumption is appropriate for all networks with different spatial and temporal resolutions. Furthermore, some alternatives, including the log-normal and stretched exponential are proven to describe real-world indication of scale-free property better than power law. The absence of scale-free property could be associated with the fitting performance of the gravity model on fine-resolution urban mobility networks. We examine the fitting performance of gravity model on fine-grained urban mobility networks in the next section.

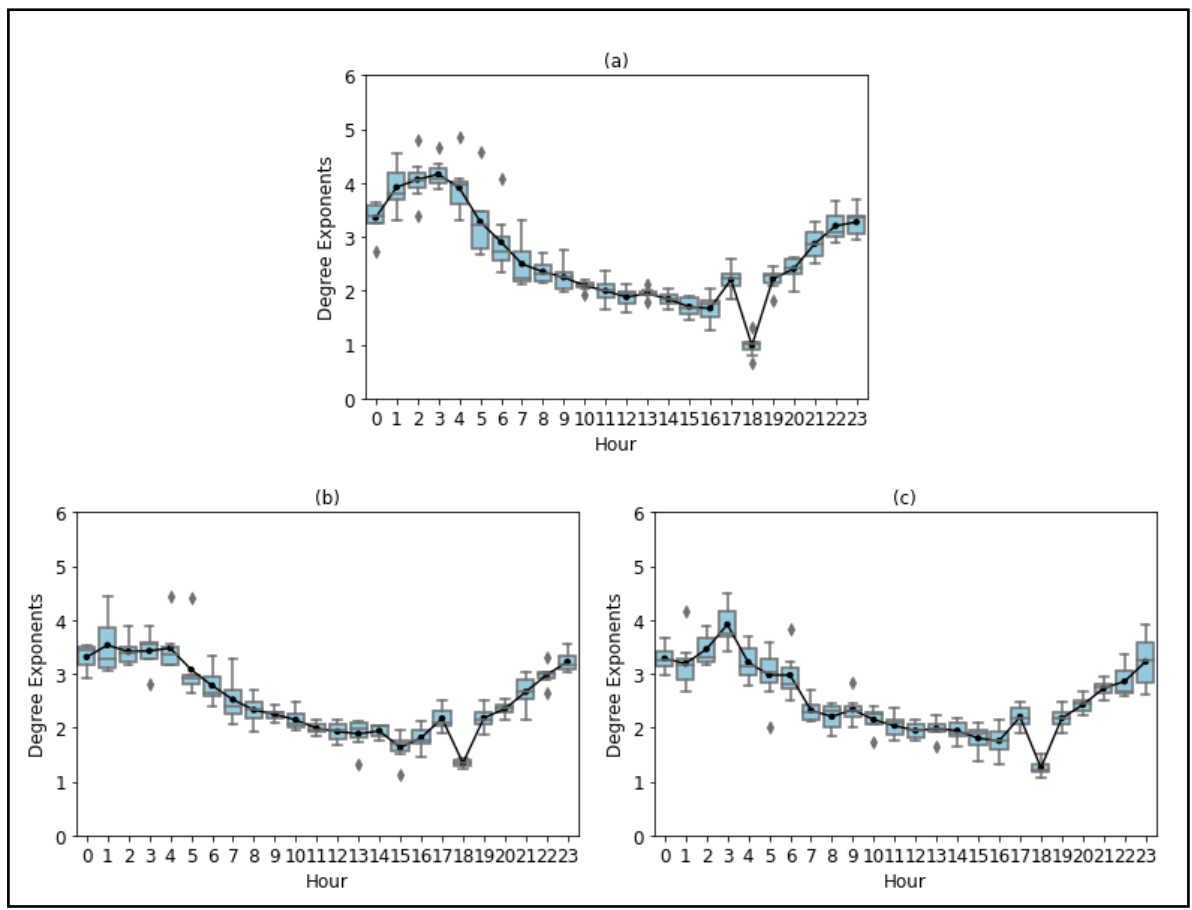

**Figure 2.** Distribution of the three types of degree exponents for the mobility network of Harris County in February 2020. (a) degree (b) in-degree (c) out-degree

## Gravity model fitting performance

To investigate how the basic gravity model performs in capturing the dynamics of spatial interaction of human movement networks using data at different resolutions, the fitting performance of the model was examined at both a daily and hourly basis. The dependent variable of the gravity model would be the flow between each origin-destination pair, while the independent variables are the population of both the origin and destination census tracts and the distance between each O-D pair. The statistical summary report consists of the coefficients and p-values of the independent variables and $R^2$ values. The coefficients and p-values indicate an independent variable's direction and extent of influence on the dependent variable. The adjusted $R^2$ values serve as an indicator of how well the model fits the variation of data. These indicators gauge the overall performance of the gravity model. The differences in the fitting results will be compared across counties and under different temporal resolutions.

**Fitting the hourly human movement network**

Figure 3 shows a general illustration of the $R^2$ values of each hour in a county during February 2020. The "-"symbol represents insufficient data. Very few or no trips are recorded for some periods due to the finer temporal resolution, at which model fitting performance values are also unavailable. For the hourly resolution, none of the independent variables is statistically significant. Their coefficients fluctuate and provide limited information about their effects on the magnitude of population flows. The model yields unstable and low $R^2$ values (<0.3) during most periods. Higher $R^2$ values at some off-peak periods do not necessarily indicate good fitting performance but may be due to their small set of outcomes and variation in terms of flow (e.g., all being 1s). Variations in $R^2$ values across counties can be observed that some have more homogeneous results during all periods while some have more significant fluctuations.

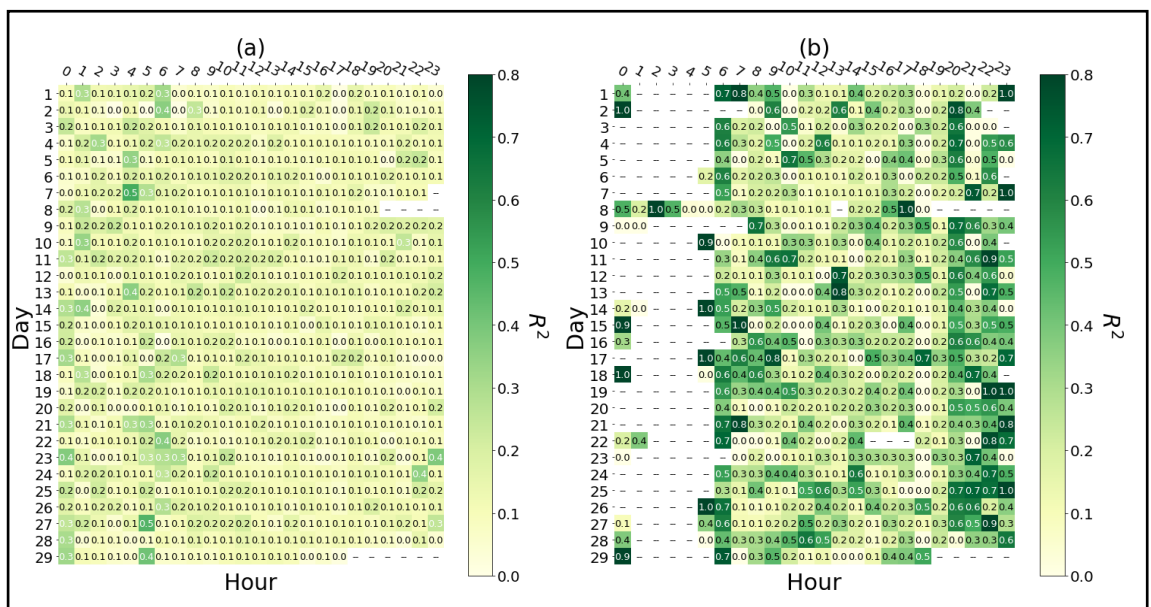

**Figure 3.** Hourly fitting performance ($R^2$) of mobility network using gravity model in February 2020. (a) Harris County (b) Queens County

**Fitting the daily human movement network**

When the temporal aggregation is daily movements, the population variables are still not significant. When the temporal aggregation is hourly movements, however, the distance variable starts to gain significance. The distance between origin and destination shows a negative effect on the flow between them. This finding is consistent with the model assumption that a shorter distance between a pair of origin and destination is supposed to form a stronger interaction. The model yields relatively higher $R^2$ values with a 0.2 to 0.6 range. Figure 4 shows the variances of $R^2$ values across the counties for each day in February 2020. There is an observable difference in the distribution of $R^2$ values between different counties, while the variance within each county is quite small. The distributions of $R^2$ values are similar each day. Those aspects of results remain stable when the data are aggregated to a coarser level, that is, to each day of a week (such as Mondays versus Wednesdays) indicating that the network structure remains about the same on the same weekdays. In the daily aggregated mobility networks, while the variances of $R^2$ values become even smaller, the overall fitting performance is not further improved. Also, the $R^2$ values are slightly higher but no significant differences were observed between different weekdays and weekends. This results indicate that even with daily aggregation, the gravity model does not yield the fitting performance similar to those reported for coarser temporal (e.g., weekly or monthly) and spatial aggregations (e.g., TAZs or ZIP codes).

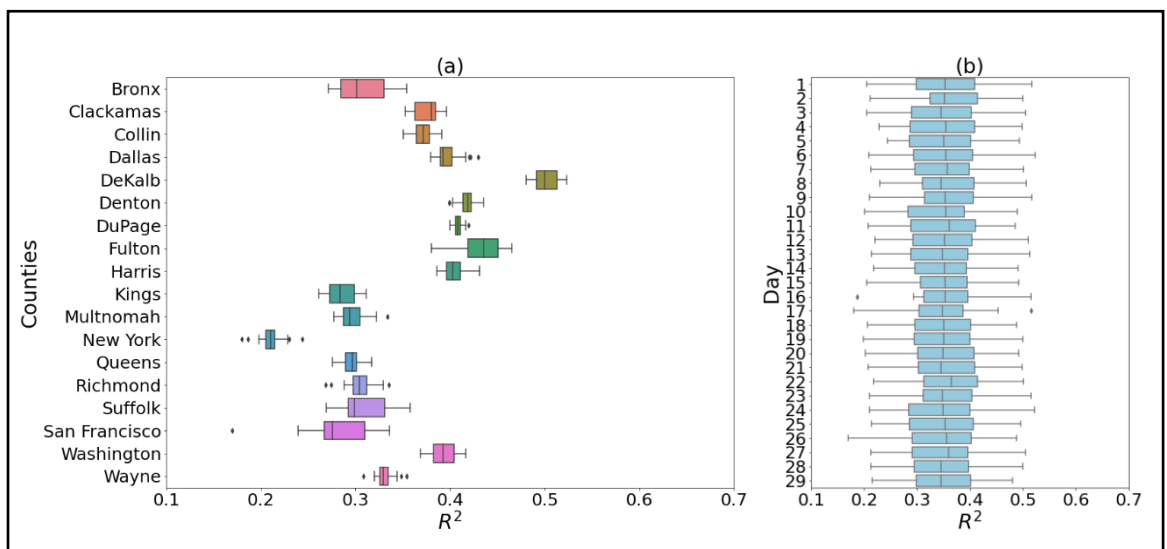

**Figure 4.** Gravity model fitting performance of mobility networks. (a) across counties (b) for different days, February 2020.

**The association between model-fitting performance and other factors**

To further examine in what conditions the gravity model performs better with a finer temporal and spatial aggregation of data, the association between fitting performance and network attributes, such as including population density of nodes and network characteristics, were evaluated. The network metrics chosen are the number of nodes and edges, average weighted degree, average clustering coefficient, average shortest path length and assortativity coefficient, which represent the change of network characteristics in terms of the size, complexity and connection. We primarily performed this analysis on the daily aggregated mobility networks.

**Variation of fitting performance with respect to population density of counties**

First, we examined how the population density of a county relates to the fitting performance of the gravity model to predict its movement network. As shown in Figure 5, both ordinary and log-scale population density is plotted against the averaged $R^2$ and the daily $R^2$ values are plotted against the population density for each county. The averaged $R^2$ values tend to be lower in counties with higher population density.

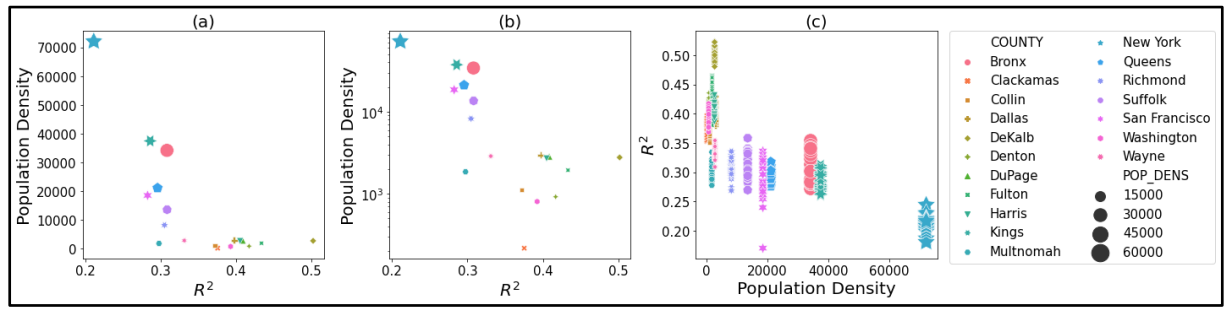

**Figure 5.** Association between fitting performance ($R^2$) of mobility networks using gravity model and population density of each county. (a) average $R^2$ – population density. (b) average $R^2$ – population density (log-scale). (c) population density – $R^2$.

Figure 6 shows population density plotted against the variances of $R^2$ values on both ordinary and log-scale. The variances of $R^2$ values are all small and the tendency that the population density is higher when the variance is higher becomes clearer.

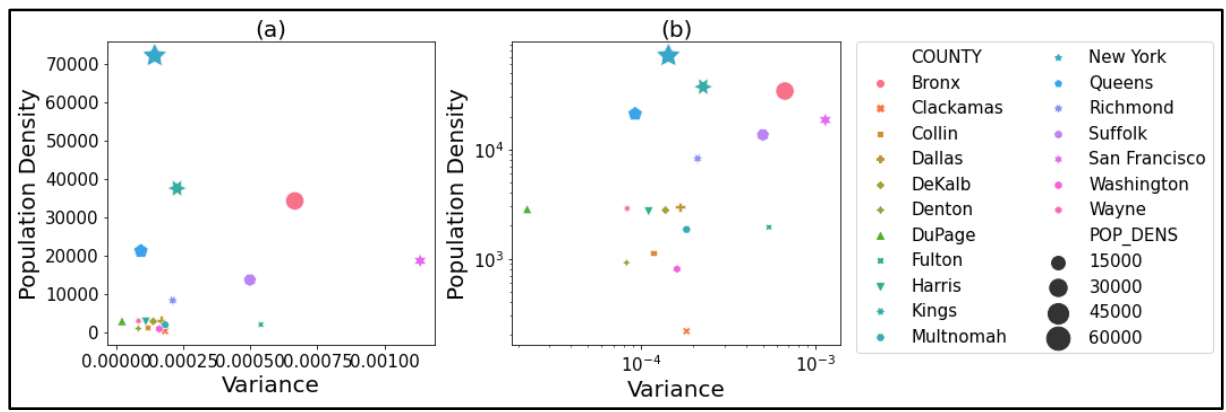

**Figure 6.** Association between the variance of fitting performance ($R^2$) of mobility networks using gravity model and population density of each county. (a) population density – variance of $R^2$. (b) population density (log-scale) – variance of $R^2$ (log-scale).

    These results show the poorer fitting performance of the gravity model for counties with higher population density. The distance between spatial blocks may be shorter in counties with higher population density. These characteristics may weaken the representativeness of the gravity model, which relies heavily on the population and distance variables. Under the same level of aggregation, counties with higher population density tend to have more data points that make their network structure denser. This may lead to poorer fitting performance because of the increase in variations of trips on links. On the other hand, there is an opposite trend for the counties with lower population density. Again, this does not necessarily mean the performance would be better in counties with lower population density, but the result may simply be due to the lower variances in the flow on the links between origin and

destinations.

**Variation of fitting performance with respect to network properties of counties**

Figure 7 displays the $R^2$ values and the average of $R^2$ values plotted against the node counts and edge counts of mobility networks in different counties. The numbers of nodes and edges of each network represent the network size. The size of the network is quite stable for daily aggregation, where the numbers of nodes and links remain almost the same for a county during different periods. Data points of every single county are distributed closely and form clusters, but there is no significant association between network size and $R^2$ values. However, there may be trends within individual counties: the more edges, the lower the $R^2$ value. This result implies that the performance of the gravity model is worse in counties with larger mobility networks; this result is consistent with the result related to counties with higher population density. When we compare network size across different counties, we realize that a small difference in the number of nodes would lead to a much greater increase in the number of links. Figure 8 illustrates the variation among the mobility network of different counties in terms of node count and edge count.

Figure 9 illustrates the $R^2$ values and their association with another set of network metrics (i.e., average weighted degree, average clustering coefficient, average shortest path length, and assortativity coefficient). This set of common network metrics helps reveal the relationship between the fitting performance and the structural properties of a network, which has larger variation over time. For $R^2$ values versus averaged weighted degrees, networks with average weighted degrees that lay on the two extremes also seem to have their fitting performances on the two extremes. The average clustering coefficients are very low across all counties. The networks with shorter average shortest path length might indicate the existence of hubs, leading to a scale-free-like structure and having better fitting performance. The assortativity coefficients of the networks for each county are all close to zero, which shows no clear segregation of nodes based on their degree values. Overall, the gravity model fitting results showed slightly better performance in some counties with smaller size and lower average distance; however, the overall fitting performance values were low across all counties compared to what was reported in previous studies for mobility networks constructed on coarser spatial and temporal scale and data aggregation.

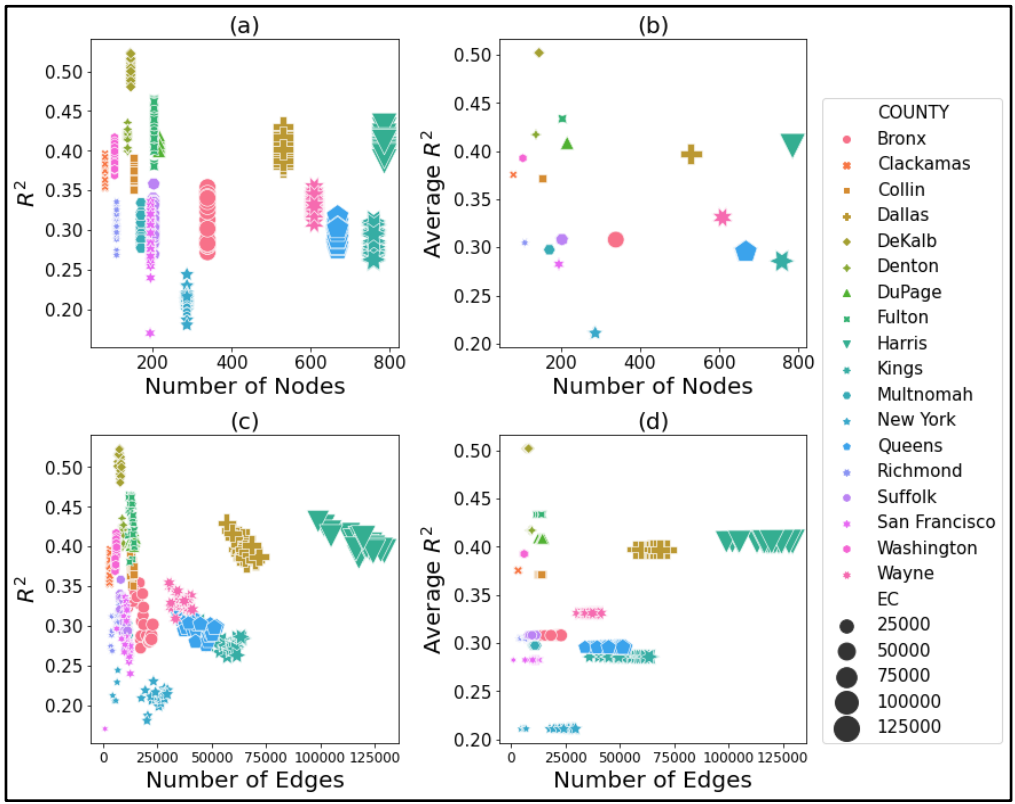

**Figure 7.** Association between network size and fitting performance ($R^2$) of the mobility networks using gravity model. (a) $R^2$ – number of nodes. (b) average $R^2$ – number of nodes. (c) $R^2$ – number of edges. (d) average $R^2$ – number of edges.

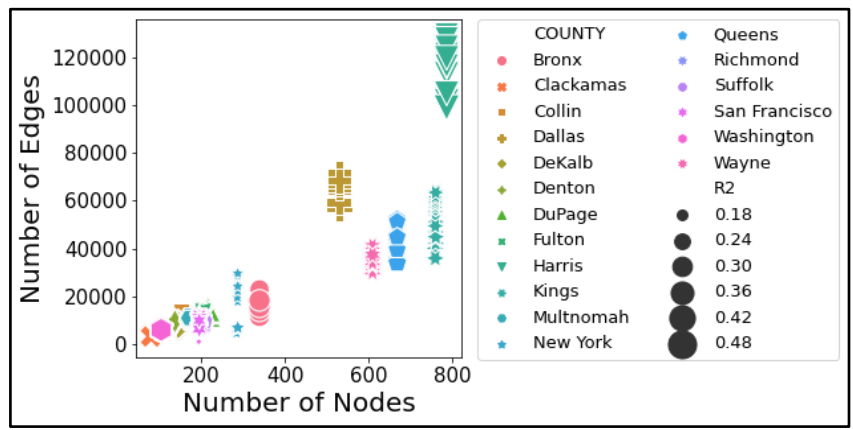

**Figure 8.** Association between the number of nodes and the number of edges within mobility networks.

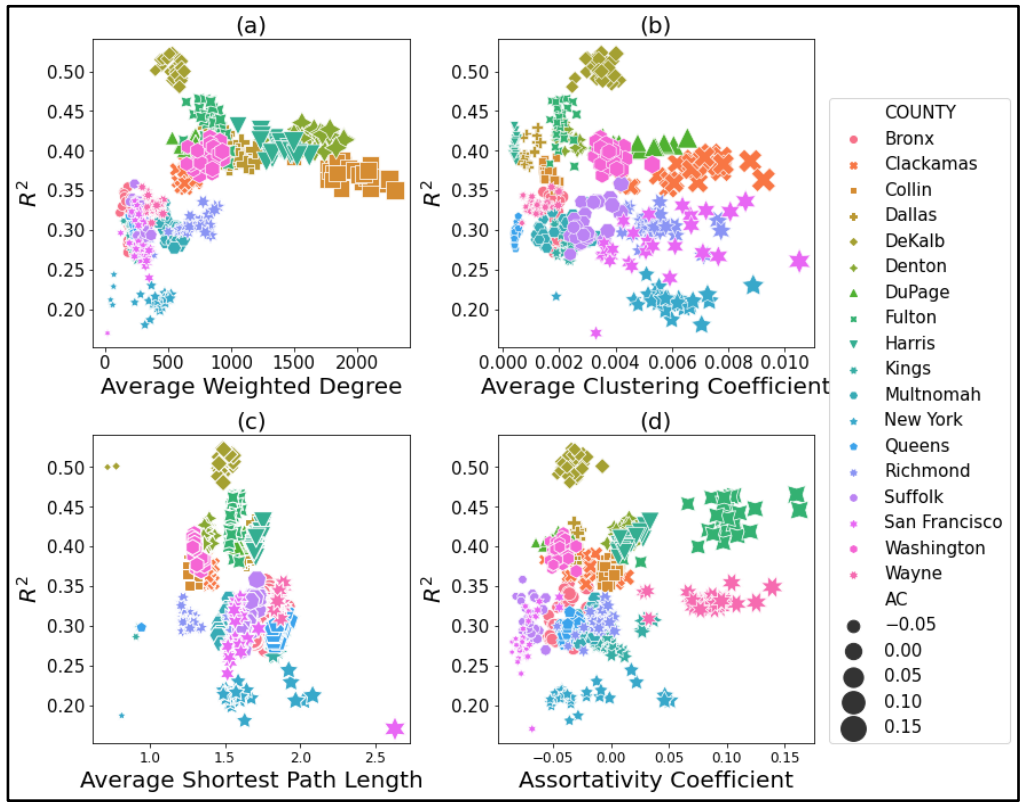

**Figure 9.** Association between the four macroscopic network characteristics and fitting performance ($R^2$) of the mobility networks using gravity model. (a) $R^2$ – average weighted degree. (b) $R^2$ – average clustering coefficient. (c) $R^2$ – average shortest path length. (d) $R^2$ – assortativity coefficient.

## Discussion

This study shows that the spatial and temporal resolution of the data source significantly affects the validity of the scale-free assumption. We compare the results and findings in our study with existing studies that use gravity models to predict the spatial interactions and population flow based on coarser spatial and temporal resolution. The hourly and daily census tract-to-census tract mobility network constructed from the location-based dataset in our study transforms our understanding of the conventional models in capturing the patterns of human mobility in cities.

While the basic gravity model does not perform well on an hourly census tract-to-census tract mobility network, the improvement of fitting performance is observed when movements were aggregated daily. The daily aggregation level could be a proper resolution that is coarse enough to record the complete network comprehensively and precise enough to capture the evolution of network structures over time. Among the three variables considered in the basic gravity model (population of the origin, population of the destination, and distance between them), the distance variable was most significant, and its negative effect on the interaction between nodes (flow) is reasonable. By checking the association of $R^2$ values with population density and network metrics, we found that network topology varies across counties but is usually consistent each day. The higher population density engenders poor fitting performance of the basic gravity model because the assumed relationship between flow, population, and distance may not be significant since the origin and destinations are usually closely located. Contrarily, mobility networks of some counties had similar network metric values but performed differently on the gravity model fitting. Data points from the same county tend to form clusters that share similar network metric values and fitting performances.

The findings in this study confer significant contributions to the underlying mechanisms of human mobility in cities. This study addressed the limitations of the basic gravity model for predicting human movement patterns in urban mobility networks at finer spatial and temporal scales and pointed out the absence of scale-free property commonly believed to be universal. The idea of the basic gravity models is to characterize the urban human movement network in a simple approach by using only the population and distance variables to analogize attraction and distraction of origin and destinations. However, because of the poor fitting performance on finer spatial and temporal resolution data, model assumptions should be examined when applied to mobility networks under finer data resolution. Insights from this study reveal the need for better-fitting models as high-resolution data become more accessible. Finer-scale urban mobility networks may violate the basic gravity model's fundamental assumptions due to their unique network structure. We could also observe that the distance variable retains a negative effect on the prediction of movement between two locations, while its significance decreases in fine-grained data. This indicates that, in reality, people's movements are the outcome of the compounding of several variables. Through the specification of time and areas, the movements would rely more on the variables, such as social connections, service needs and work requirements, in addition to distance. Our findings inform a diverse and great amount of new research directions for further exploration of the nature of human mobility with other dimensions. Future studies could focus on the following directions to obtain better modeling results: (1) Identify the attributes or develop metrics that could better describe the urban mobility network to be included in the models. (2) Clarify the limitations of the use of the existing models on mobility networks and pinpoint adjustments necessary to enhance performance. (3) Introduce new approaches, such as machine learning-based approaches, to account for other features of the finer spatial blocks (such as their points of interest). In addition, prior studies (Alessandretti et al., 2020) have identified the effects of scale on the frequency of movements. Our study is in response to this prior study, contributing to the understanding of mobility at the census-tract level and the limited predictability of gravity models at this scale. Future research could look into finer scales, such as the neighborhood level or block level and specify other influential factors on human mobility. Our study opens a door to examine and reveal more dimensions motivating people's movements. The integration of all findings with fine-grained data related to movements will advance and transform prior knowledge of human mobility in cities.

Existing studies focus on adopting and developing mathematical or statistical models to explain the patterns of human mobility. These models, however, usually are not able to integrate the variables and factors from distinct aspects. As the science of artificial intelligence (AI) techniques advances (Luca et al., 2021; Toch et al., 2019; Wang et al., 2019; Zhao et al., 2016), more research works have demonstrated the performance of AI in capturing the complex relationships among factors in different dimensions. For example, trajectory generative adversarial network (TGAN) can use motion patterns and location distribution to eventually identify human mobility (Zhou et al., 2021). DeepMove, an attentional recurrent network for mobility prediction from lengthy and sparse trajectories (Feng et al., 2018) could [describe DeepMove abilities]. With the use of AI and fine-grained data, future research could devote to developing AI-based models to understand and predict human mobility in cities.

## Concluding remarks

In this study, we first constructed the human mobility network based on fine spatial-temporal resolution data. Second, the commonly assumed scale-free property was examined in these finer-scale urban mobility networks. Third, the validity of the basic gravity model in predicting human mobility under daily and hourly aggregations and at the census-tract level was tested and compared to the results in other studies. Then, we examined the association of network metrics and population density with the performance of the gravity model across different counties. The results show that the scale-free property does not always exist in real-world human mobility networks; in fact, it is very rare. The fitting performances of the basic gravity model on both hourly and daily human mobility network for the census tract-to-census tract movements are worse than the fitting performance values reported on coarser aggregated data. From these results, we can conclude some key findings regarding the limitations of the application of basic gravity models for modeling spatial-temporal mobility networks in urban areas. The basic gravity model is not suitable for modeling human mobility networks at finer spatial and temporal scales. Among the independent variables used in the basic gravity model, the distance variable may the most significant. This results indicate that population density of census tracts does not play a key role in population flow across census tracts. Other features (such as presence of points of interest and land use features) might be more crucial to drive the population flow across census tracts.

## Acknowledgments

This material is based in part upon work supported by the National Science Foundation Critical Resilient Interdependent Infrastructure Systems and Processes (CRISP) program, Texas A&M University X-Grants program, and Microsoft Azure. The authors also would like to acknowledge the data support from X-Mode. Any opinions, findings and conclusions, or recommendations expressed in this material are those of the authors and do not necessarily reflect the views of the National Science Foundation, Texas A&M University, X-Mode, or Microsoft Azure.

## Data availability

The data that support the findings of this study are available from X-Mode, but restrictions apply to the availability of these data, which were used under license for the current study. The data can be accessed upon request submitted on X-Mode. Other data used in this study are all publicly available.

## Code availability

The code that supports the findings of this study is available from the corresponding author upon request.